\documentclass{article}
\usepackage{spconf,amsmath,graphicx,hyperref}
\usepackage{enumitem,amssymb, multirow}

\title{An Event-based Sequence Modeling Approach \\to Recognizing Non-Triad Chords with Oversegmentation Minimization}
\name{Leekyung Kim$^{1}$, Jonghun Park$^{1}$}
\address{$^{1}$Department of Industrial Engineering and Institute of Industrial Systems Innovation,\\
Seoul National University, Seoul, Republic of Korea}

\begin{document}
%
\maketitle
\begin{abstract}
Automatic chord recognition (ACR) extracts time-aligned chord labels from music audio recordings. Despite recent advances, ACR still struggles with oversegmentation, data scarcity, and imbalance, especially in recognizing complex chords such as non-triads, which are unpopular in existing datasets. To address these challenges, we reformulate ACR as a segment-level sequence-to-sequence prediction task, where chord sequences are predicted auto-regressively rather than frame by frame. This design mitigates excessive segmentation by detecting chord changes only at segment boundaries. We further introduce two types of token representations and an encoder pre-training method, both specifically designed for time-aligned chord modeling. Experimental results show that our model improves performance in both chord recognition and segmentation, with notable gains for complex and infrequent chord types. These findings demonstrate the effectiveness of segment-level sequence modeling, structured tokenization, and representation learning for advancing chord recognition systems.
\end{abstract}
\section{Introduction}
\label{sec:intro}
\begin{figure}[t]
    \includegraphics[width=\linewidth]{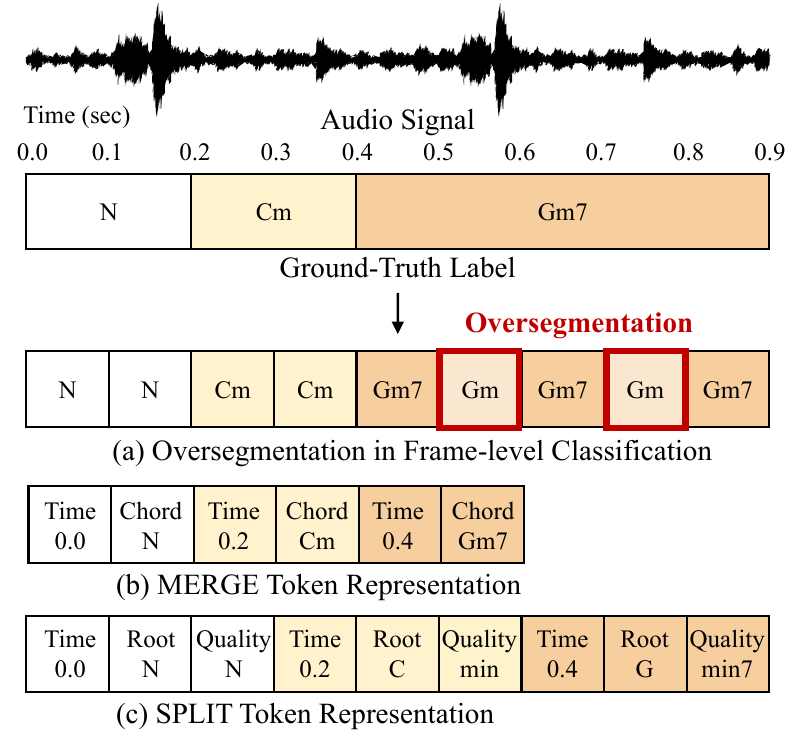}
    \centering
        \caption{Comparison of (a) frame-level classification and the proposed token representations: (b) MERGE and (c) SPLIT.}
    \label{fig_token}
\end{figure}

Automatic Chord Recognition (ACR) is the task of identifying the sequence of chords in musical audio, where chords---a group of simultaneously played notes---form the harmonic foundation of a musical piece\cite{20year}. ACR remains an active area of research due to complex musical structures and diverse instrumentation, which make chord annotation time-consuming and demand highly specialized musical expertise.

Transformer\cite{transformer}-based models have recently achieved state-of-the-art performance in ACR\cite{btc,ht,atc}, leveraging their success in natural language processing\cite{bert}. BTC\cite{btc} employed an encoder-only Transformer for audio input, while Harmony Transformer (HT)\cite{ht,atc} adopted a full encoder-decoder structure, performing segmentation with the encoder and recognition with the decoder on symbolic input. In this work, we focus on ACR from audio, which is more applicable to real-world music scenarios.

Recent studies have expanded the target vocabulary from 25 basic chords to a broader range of chord types\cite{gaussian_hmm,decomposition}, including non-triad chords. However, such complex chords are unpopular in datasets, leading to data imbalance and limited generalization. To address these issues, prior works have explored structured chord representations\cite{decomposition, structured}, even chance training\cite{evenchance}, curriculum learning\cite{cl}, semi-supervised learning \cite{unl, contrastive}, and adaptive loss functions\cite{unl}.

Despite these advances, deep learning-based ACR methods still face three major challenges. (1) Oversegmentation: In traditional frame-level classification, probabilities are predicted at every time frame, making them highly sensitive to small fluctuations. A single misclassified frame can split a chord segment and cause excessively fragmented and unstable boundaries\cite{decomposition}, as shown in Fig. \ref{fig_token}(a). By contrast, our approach predicts chords at segment level, detecting chord changes only when they occur, thereby mitigating oversegmentation. To this end, we reformulate ACR as a segment-level sequence-to-sequence (seq2seq) learning problem and introduce a Transformer encoder-decoder that predicts chord sequences auto-regressively.

(2) Data scarcity: The difficulty of chord annotation and copyright restrictions limit the availability of large-scale labeled datasets\cite{chord_ds}. (3) Data imbalance: The skewed chord distribution biases models toward common chords, while underrepresenting unpopular or complex ones\cite{chord_di}. To address these data-related issues, we propose two types of token representations---MERGE and SPLIT---and an encoder pre-training method based on chord similarity.

The contributions of this work are as follows:
\begin{itemize}[noitemsep]
\item We reformulate ACR as a segment-level seq2seq task and propose an auto-regressive chord prediction framework using a Transformer encoder-decoder architecture. To the best of our knowledge, this is the first study to apply this approach to ACR.
\item We introduce two token representations for time-aligned chord modeling and a chord similarity-based encoder pre-training strategy to address data scarcity and imbalance.
\item We conduct experiments demonstrating that our model\footnote{\url{https://github.com/KimLeekyung/ACR_seq2seq}} achieves state-of-the-art performance in both chord recognition and segmentation.
\end{itemize}

\section{Proposed Method}

\subsection{Problem Specification}
Most traditional ACR models predict the probability of chords at each time frame from an input spectrogram $X_{\text{spec}}\in \mathbb{R}^{N_T\times N_F}$, where $N_T$ is the number of time frames and $N_F$ is the number of frequency bins. 
This can be formulated as ${\hat {y}_{i}} = \arg\max_{y_i\in\mathcal{V}}P(y_{i}\vert X_{{\text{spec}}}) \mbox{ for } i=1,\dots,N_T $, 
where $\mathcal{V}$ is a pre-defined chord vocabulary, and $\hat{y}_i$ is the predicted chord for the $i^{th}$ time frame, selected as the class with the highest probability among $y_i\in\mathcal{V}$.

This paper models ACR as a seq2seq problem that auto-regressively predicts token sequence from audio sequence, which is formulated as ${\hat {\sigma}_{j}} = \arg\max_{\sigma_j\in{\Sigma}}P(\sigma_{j}\vert X_{{\text{spec}}}, \sigma_{1:j-1})$ $\mbox{for } j=2,\dots$,
where $\sigma_j$ is the $j^{th}$ token in a token sequence and $\Sigma$ is a fixed token set. A predicted token $\hat{\sigma}_j$ is the one with the highest probability, given the spectrogram and the previous token sequence $\sigma_{1:j-1}$. The first token, $\sigma_1$, is fixed as the start-of-sequence token, and the token prediction continues until the end-of-sequence token is predicted or the token sequence reaches the maximum length.

\subsection{Model Structure and Chord Representation}
\begin{figure}[!t]
    \includegraphics[width=0.8\linewidth]{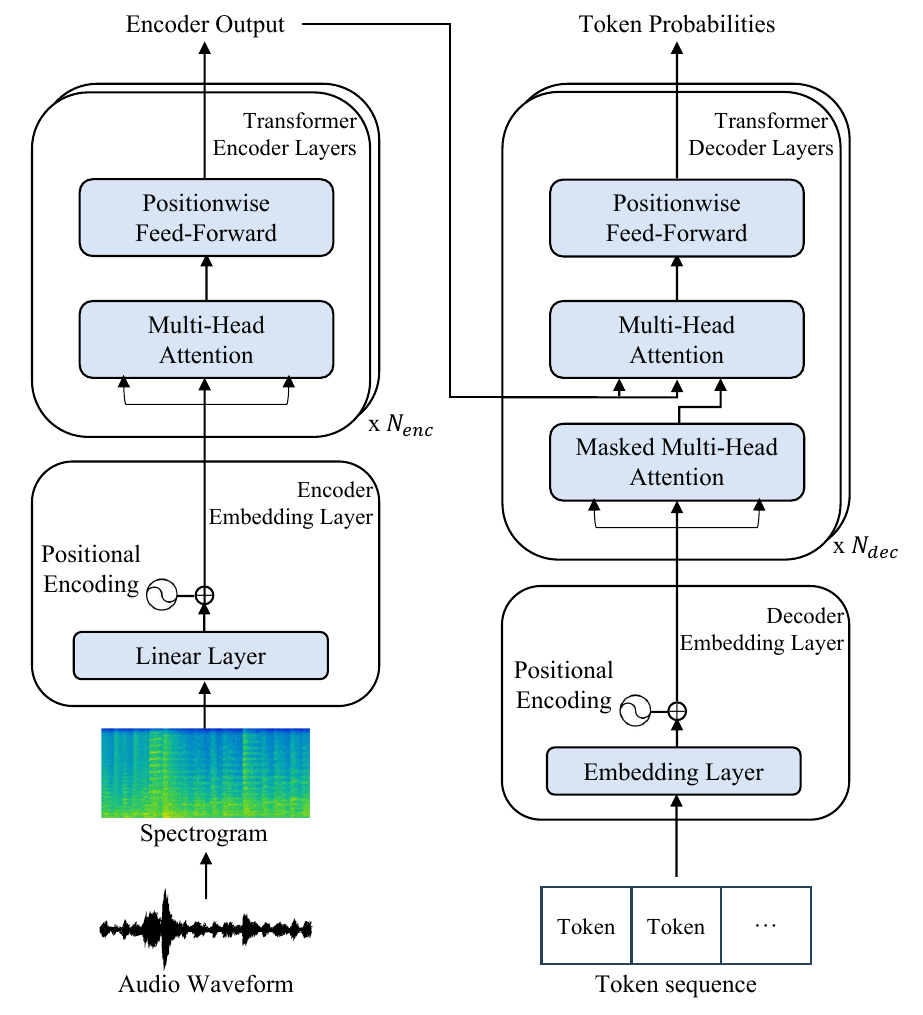}
    \centering
    \caption{Overall architecture of the Transformer encoder-decoder model. $N_{enc}$ and $N_{dec}$ denote the numbers of encoder and decoder layers, respectively.}
    \label{fig_model}
\end{figure}

This paper adopts the Transformer encoder-decoder structure due to its improved performance and suitability for auto-regressive sequence prediction, as illustrated in Fig. \ref{fig_model}. Audio segments of $T$ seconds are first converted into waveforms and then transformed into spectrograms. To reduce excessive segmentation\cite{korzeniowski2017futility} and eliminate the need for additional modules such as conditional random fields\cite{crf} often required to smooth predictions in the traditional methods\cite{decomposition, korz2016cnn}, we propose two types of token representations, MERGE and SPLIT, illustrated in Fig. \ref{fig_token}(b) and Fig. \ref{fig_token}(c), respectively.

In MERGE, each time-aligned chord is expressed using a time token indicating the chord boundary and a chord token representing the full chord from $\mathcal{V}$. As $\mathcal{V}$ includes `no chord' and `unknown chord', chord durations are implicitly defined as continuing until the next chord onset. SPLIT decomposes each chord token into root and quality tokens. For instance, \textit{C:maj} is divided into the root \textit{C} indicating the chord's fundamental note, and the quality \textit{maj}, describing its interval structure. This decomposition enables multiple chords, such as \textit{C:dim} and \textit{B:dim}, to share the same quality token, \textit{dim}. As a result, SPLIT increases the amount of training data available for unpopular chord qualities and facilitates the learning of structural similarities across chords, thereby improving the robustness of learning for unpopular chord qualities.

The token sets are defined as follows: 
$\Sigma_{\text{M}} = \mathbb{T} \cup \mathbb{C} \cup \mathbb{S}$ for MERGE, and $\Sigma_{\text{S}} = \mathbb{T} \cup \mathbb{R} \cup \mathbb{Q} \cup \mathbb{S}$ for SPLIT, where $\mathbb{T}$, $\mathbb{C}$, $\mathbb{R}$, $\mathbb{Q}$, and $\mathbb{S}$ denote the sets of time, chord, root, quality, and special tokens, respectively. The time token specifies absolute time within each segment at 0.1-second intervals. The chord token in MERGE represents each chord from $\mathcal{V}$. The root token includes all 12 pitch classes and a shared token for `no chord' and `unknown chord'. The quality token spans all chord qualities in $\mathcal{V}$, with separate tokens for `no chord' and `unknown chord'.
The special tokens \texttt{<SOS>}, \texttt{<EOS>}, and \texttt{<PAD>} represent the start, end, and padding of sequence, respectively.

\subsection{Encoder Pre-Training}

We hypothesize that an effective encoder improves ACR performance by producing similar embeddings for audio segments with similar chord sequences. Based on this, we introduce an encoder pre-training method to learn meaningful embeddings guided by chord similarity. The encoder is first pre-trained and later used to initialize the encoder weights during full model training, while the decoder is randomly initialized.

We adopt the Weighted Chord Symbol Recall (WCSR), one of the most widely used metrics for evaluating chord recognition performance, to compute chord similarity. It measures the similarity between two chord sequences by computing the ratio of correctly matched time frames based on harmonic content\cite{metric-harte}. Among various matching criteria with different levels of strictness, we select the \texttt{mirex} criterion, which considers two chords matched if they share at least three pitch classes, as it offers a balanced evaluation.

During pre-training, a comparison sample $X_2$ is randomly selected from the batch for each anchor spectrogram $X_1$, excluding the anchor itself. Chord similarity is computed from the corresponding chord labels $Y_1$ and $Y_2$. The encoder embeddings $o_1$ and $o_2$ are compared using cosine similarity, which shares a similar range and interpretation with WCSR. Only the encoder is trained to minimize the mean squared error (MSE) between the embedding similarity and the ground-truth chord sequence similarity.

\section{Experiments}
\subsection{Experimental Setups}
\textbf{Data.} We use the dataset as BTC\cite{btc}, consisting of 471 pop songs with manually aligned audio and chord annotations. A 5-fold cross-validation was conducted by splitting the whole data into five subsets. In each fold, one subset was used for validation and the remaining four subsets for training, applied identically to both pre-training and full model training. 

\noindent\textbf{Implementation Details.}
Audio files were divided into 25.6-second segments, sampled at 44,100 Hz, and converted to constant-Q transform spectrograms\cite{cqt} with a hop length of 4,410, spanning 6 octaves from C1 at 24 bins per octave, using log amplitude scaling.
Following the previous ACR research\cite{btc,cl,contrastive}, $\mathcal{V}$ includes `no chord', `unknown chord', and 168 chords derived from 12 pitch classes combined with 14 chord qualities: \textit{maj}, \textit{min}, \textit{dim}, \textit{aug}, \textit{min6}, \textit{maj6}, \textit{min7}, \textit{minmaj7}, \textit{maj7}, \textit{7}, \textit{dim7}, \textit{hdim7}, \textit{sus2}, \textit{sus4}. Accordingly, there are 257 time tokens, 170 chord tokens, 13 root tokens, 16 quality tokens, and 3 special tokens, resulting in 430 tokens for $\Sigma_M$ and 289 tokens for $\Sigma_S$. 

\noindent\textbf{Training Details.}
The encoder was pre-trained using MSE loss, and the full model was trained with cross-entropy loss. Both stages used the Adam\cite{adam} optimizer, with the learning rate halved if validation loss did not decrease for 3 epochs. Early stopping was applied for 10 stagnant epochs. 
Pitch shifting and random cropping were employed for data augmentation on root and time tokens, respectively.

\noindent\textbf{Inference.}
At each step, the token with the highest probability is selected, and masking is applied to enforce the pre-defined token order, such as (time token, chord token) for MERGE. As summarized in Table \ref{masking_tab}, `O' and `X' indicate whether a token type is allowed or masked, respectively, based on the previously predicted token. `-' denotes that the token type is not used in the corresponding representation.

\begin{table}[]
\centering
\caption{Masking strategy for inference.}
\resizebox{\linewidth}{!}{%
\begin{tabular}{c|c|cccc}
\hline
\multirow{2}{*}{\begin{tabular}[c]{@{}c@{}}Token\\ Representation\end{tabular}} & \multirow{2}{*}{\begin{tabular}[c]{@{}c@{}}Previous\\ Token Type\end{tabular}} & \multicolumn{4}{c}{Next Token Type} \\
                                                                                &                                                                                & Time   & Chord   & Root  & Quality  \\ \hline
\multirow{2}{*}{MERGE}                                                          & Time                                                                           & O      & X       & -     & -        \\
                                                                                & Chord                                                                          & X      & O       & -     & -        \\ \hline
\multirow{3}{*}{SPLIT}                                                          & Time                                                                           & O      & -       & X     & O        \\
                                                                                & Root                                                                           & O      & -       & O     & X        \\
                                                                                & Quality                                                                        & X      & -       & O     & O        \\ \hline
\end{tabular}
}
\label{masking_tab}
\end{table}

\noindent\textbf{Metrics.}
WCSR was calculated with seven criteria (see Table \ref{wcsr_tab}) to assess recognition performance at varying levels of strictness. Segmentation quality (SQ) metric, measured with the directional Hamming distance (DHD)\cite{dhd-mau}, was used to evaluate segmentation performance. 
DHD is computed by identifying the maximum overlapping segments between two sequences and summing their differences. It evaluates over- and undersegmentation, and SQ is derived from these two measures to evaluate the overall segmentation quality\cite{metric-harte}. 

\subsection{Quantitative Results}

\begin{table*}[!ht]
\centering
\caption{WCSR and SQ scores for experiments.}
\begin{tabular}{l|ccccccc|ccc}
\hline
\multicolumn{1}{c|}{\multirow{2}{*}{}} & \multicolumn{7}{c|}{WCSR}                                                                                                                                                                                  & \multicolumn{3}{c}{SQ}                        \\
\multicolumn{1}{c|}{}                  & \multicolumn{1}{c}{root} & \multicolumn{1}{c}{maj-min} & \multicolumn{1}{c}{thirds} & \multicolumn{1}{c}{triads} & \multicolumn{1}{c}{sevenths} & \multicolumn{1}{c}{tetrads} & \multicolumn{1}{c|}{mirex} & under         & over          & mean          \\ \hline
TE                                     & 81.5                     & 81.0                        & 79.6                       & 75.5                       & 71.8                         & 66.1                        & 79.6                       & 89.5          & 81.4          & 80.3          \\
TE-DM                                  & 85.6                     & 84.7                        & 83.8                       & 79.6                       & 75.7                         & 70.4                        & 83.9                       & 88.6          & 92.4          & 87.4          \\
TE-DS                                  & 86.5                     & 85.6                        & 84.9                       & 80.6                       & 77.1                         & 72.0                        & 84.9                       & 89.3          & 92.3          & 88.0          \\
pTE-DS                                 & \textbf{87.4}            & \textbf{86.7}               & \textbf{85.9}              & \textbf{81.5}              & \textbf{78.6}                & \textbf{73.2}               & \textbf{85.7}              & 89.8          & \textbf{92.9} & \textbf{88.6} \\ \hline
BTC                                    & 83.5                     & 82.3                        & 80.8                       & 75.9                       & 71.8                         & 65.5                        & 80.8                       & \textbf{90.1} & 85.9          & 84.6         \\ \hline
\end{tabular}
\label{wcsr_tab}
\end{table*}

The WCSR and SQ scores, averaged over 5 folds, are summarized in Table \ref{wcsr_tab}. We conducted an ablation study with four models to assess the contributions of each component: TE, an encoder-only model for frame-level classification; TE-DM, an encoder-decoder with MERGE; TE-DS, an encoder-decoder with SPLIT; and pTE-DS, TE-DS with encoder pre-training. We employed BTC\cite{btc}, a state-of-the-art model for end-to-end chord recognition from audio, as a comparison model. As SQ scores for BTC were not reported, we computed them using its publicly available checkpoint\footnote{https://github.com/jayg996/BTC-ISMIR19}.

Across all WCSR criteria, chord recognition performance improved consistently as each component was added to the baseline model, TE. The final model, pTE-DS, outperformed BTC.  
Moreover, as the evaluation criteria became stricter (from \texttt{root} to \texttt{tetrads}), the performance gap between pTE-DS and BTC widened, even as absolute scores decreased. This indicates the effectiveness of our tokenization and representation learning in recognizing complex chords. For segmentation, TE-DM and TE-DS achieved higher SQ than TE, demonstrating that the encoder-decoder structure with auto-regressive prediction enhances segmentation quality. pTE-DS also outperformed BTC in SQ, particularly in over-segmentation, confirming the benefits of seq2seq modeling for segmentation performance.

\subsection{Qualitative Analysis}
\begin{figure}[t]
    \centering
    \includegraphics[width=0.85\linewidth]{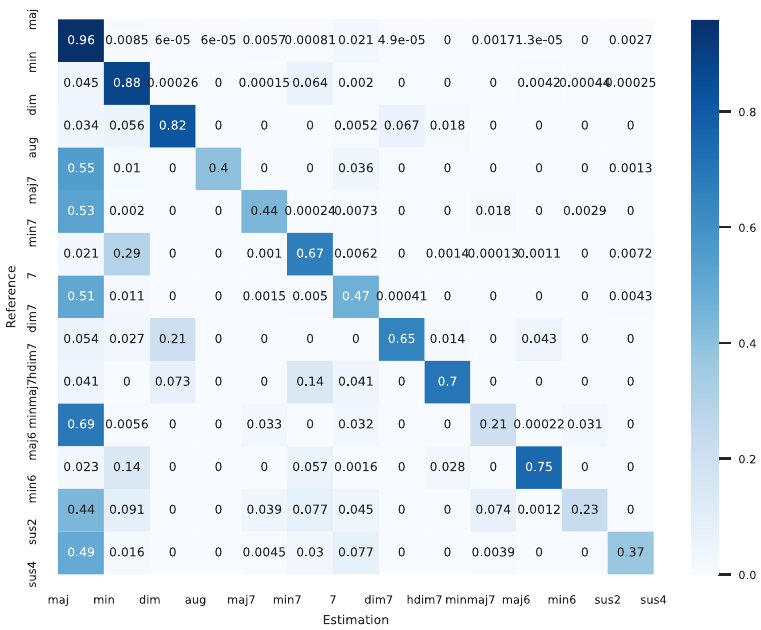}
\caption{Confusion matrix of pTE-DS with respect to quality when root notes of the prediction and reference are same. The value at row $i$ and column $j$ represents the ratio of frames labeled as quality $i$ but predicted as quality $j$.}
    \label{fig_confusion_matrix}
\end{figure}

\begin{figure}[t]
    \centering
    \includegraphics[width=0.7\linewidth]{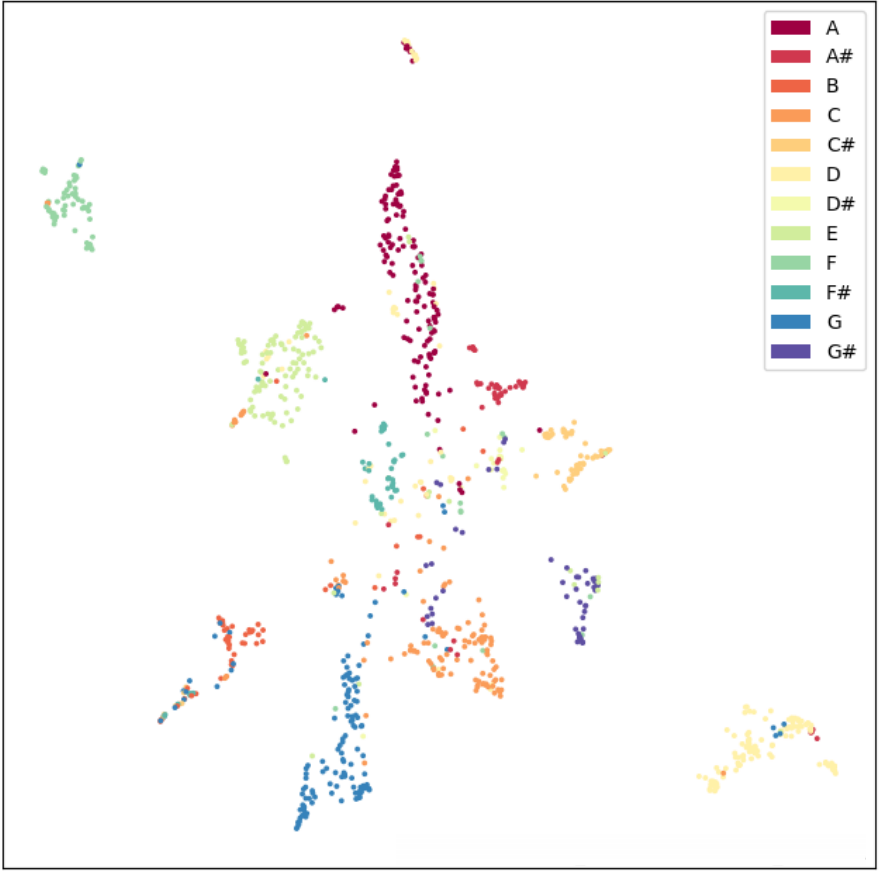}
\caption{UMAP visualization of the encoder embeddings.}
    \label{fig_umap}
\end{figure}

Fig. \ref{fig_confusion_matrix} presents the frame-wise confusion matrix for pTE-DS. While recognition of complex chords has improved, some misclassifications remain. These errors often involve simplifying complex chords to more common ones. For example, `\textit{maj6}' was frequently predicted as `\textit{maj}', which differs by one pitch class. This suggests not only a bias toward frequent chord qualities but also a difficulty in distinguishing chord qualities that are hierarchically related.

To further analyze the encoder pre-training, we visualized the learned latent space using uniform manifold approximation and projection (UMAP)\cite{umap}. 
We randomly selected 10 songs not used in training and segmented them into clips where a single chord lasted for at least one second. 
The encoder embeddings for these segments were projected into a 2D space using UMAP. 
Fig. \ref{fig_umap} shows the result, where each point represents a chord in an audio segment and colors indicate the root note of chords. The visualization reveals that the embeddings are well-clustered by the root note, indicating that the encoder successfully captures chord-related harmonic information from the input audio.

\section{Conclusion}
In this paper, we proposed a seq2seq approach to automatic chord recognition using a Transformer encoder-decoder. Our method carried out segment-level auto-regressive prediction, which alleviates oversegmentation, in contrast to traditional frame-level classification. We also introduced two types of token representations for time-aligned chords and an encoder pre-training scheme guided by chord similarity, enabling the encoder to learn musically meaningful latent embeddings. Experimental results demonstrated that our approach significantly improved both chord recognition, particularly for unpopular and complex chords, and chord segmentation performance. Qualitative analyses further revealed common error patterns and confirmed the encoder embeddings capture chord-related information effectively. Future work includes expanding recognition to a larger chord vocabulary and addressing the inherent subjectivity of chord annotations.

\newpage
\bibliographystyle{IEEEbib}
\bibliography{mainbib}

\end{document}